# Electrostatic and hydrophobic patches on Amyloid-β oligomers govern their fractal self-assembly: Implication to proteins

Anurag Singh[a], Suparna Khatun[a], and Amar Nath Gupta[a,*]

[a]Biophysics and Soft Matter Laboratory, Department of Physics, Indian Institute of Technology Kharagpur-721302, India

*Corresponding author Email: ang@phy.iitkgp.ac.in

## Abstract

We present a generic model to describe the fractal self-assembly of proteins in terms of electrostatic and hydrophobic interactions. The predictions of the model were correlated with the simulated fractals obtained using patchy diffusion-limited aggregation, and the experimentally observed Amyloid-β (Aβ) fractal self-assembly using confocal microscopy. The molecular docking was used to determine the properties of the patches on Aβ oligomers. Similar patch properties were used to design the particles for simulation. In agreement with the model predictions, the free energy of formation for the simulated fractal self-assembly was proportional to its fractal dimension; moreover, their morphologies were similar to the fractal morphologies of Aβ observed experimentally at different pH.

**Keywords:** Protein, Fractal, self-assembly, Electrostatic and hydrophobic patches, Amyloid-beta (Aβ), Molecular docking, Patchy DLA, Microscopy

The complete understanding of protein aggregation under different physiological conditions is indispensable to address many unanswered questions in protein biophysics. Notably, the misfolding of intrinsically disordered proteins (IDPs) has been linked to various neurodegenerative diseases[1]. Among these IDPs, the most studied is the amyloid-β (Aβ), which is believed to be associated with Alzheimer's disease[2]. The possibility of an enormous number of distinct polymorphs of Aβ[3] formed during the conversion of misfolded monomers into well-defined structural units called fibrils has fascinated researchers across disciplines[4]. This is the reason why most of the in-vitro studies being conducted are aimed to elucidate the aggregation kinetics of proteins; hence are mostly performed in bulk. The small size oligomers formed in the process are implicated in cell toxicity[5]. These oligomers have electrostatic and hydrophobic regions on their surface which interact with the membrane to cause its disruption[6,7].

However, when a protein solution is put under diffusion limited-conditions, for instance, when drop cast on a substrate, it may lead to its fractal-assembly[8-12]. The fractal self-assembly (FSA) of structurally different proteins has been widely observed[8-12]. The previous reports have demonstrated high sensitivity of the fractal morphology on the nature of the substrate, pH, temperature and ionic strength of the media. Also, the recent studies indicate that a sufficient number of specifically sized oligomers are necessary to observe the FSA of protein[9, 10]. However, the physicochemical interactions which lead to this FSA are still not clearly understood. Herein, we present a model to get insights into the FSA of proteins, applied particularly on the FSA obtained from the freshly prepared solution of Aβ.

In a drop-casted protein solution, the formation of FSA is preceded by usual protein aggregation[13]. The electrostatic interaction[14] provides the appropriate Gibb's free energy for a sufficient number of monomers in the misfolded state to form oligomers (Stage 1, Fig. 1(a)). Due to the confined environment in a drop cast on a substrate, the translational degree of freedom of the formed oligomers is restricted in the $-z$ direction and is partly lost in the $+z$ direction[15, 16] (Stage 2, Fig. 1(b)). This leads to an effective diffusion of the oligomers in the $x - y$ plane on the substrate. The rate limiting step in the diffusive process is the attachment of oligomers to other oligomers, which act as the nucleation site on the substrate. Therefore, provided oligomers are not stable enough, they can self-assemble into fractal-like morphologies via diffusion-limited aggregation, where the sticking probability is governed by the electrostatic and hydrophobic patches present on the solvent-accessible surface area (SASA) of the protein oligomers. The oligomers which approach the self-assembly from the $+z$ axis is responsible for the finite thickness of the self-assembly along the $z$ direction.

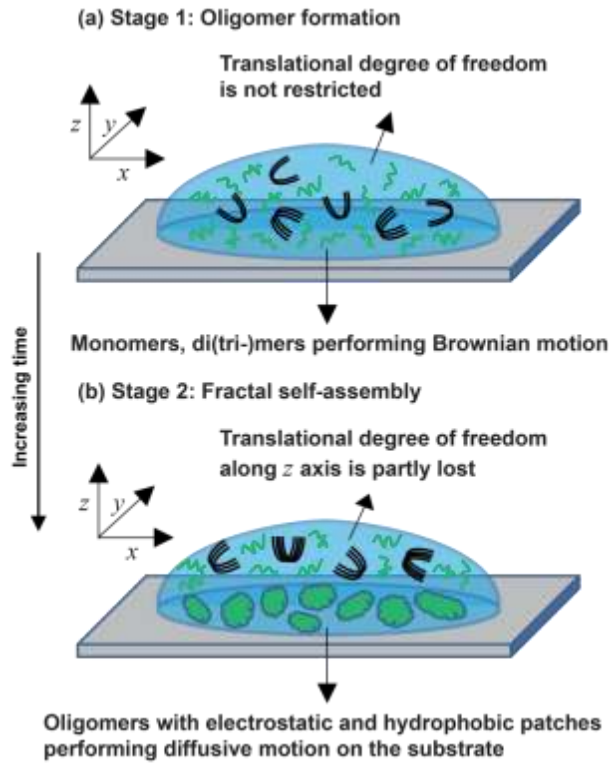

Fig. 1 Schematic representation of stages involved during the FSA of Aβ when drop cast on a substrate. (a) Freshly prepared Aβ solution contains majorly the monomers, dimers, and trimers which perform Brownian motion. (b) Later on, with an increase in time, the translational degree of freedom of the formed oligomers is restricted in the -$z$ direction and is partially lost in the +$z$ direction due to which the oligomers, having electrostatic and hydrophobic patches of their SASA, perform diffusive motion on the substrate.

To understand the FSA of protein, we present a model based on two important observations. The first result is from our recent light scattering study conducted at pH 6.5±0.1 on human amylin, a 37 residue peptide associated with type-II diabetes mellitus[10]. When the matured fibrils of human amylin, which themselves did not form FSA, were sonicated (at ultrasonic power of 250 W for 10 min and 30 min) which leads to an observation of FSA. The sonication of matured fibrils breaks them into oligomers and small protofilaments, which then diffuse on the substrate to self-assemble into fractal-like morphologies. The study showed that a sufficient number of specifically sized oligomers are required for the formation of FSA. Similar observations were made for matured fibrils of Aβ.

The second result we obtained is from the molecular docking study where mainly the electrostatic and hydrophobic interactions that may leads to FSA. The docking study provides the position and the fractional coverage of electrostatic and hydrophobic patches on the oligomers. The Cluspro webserver[17] used for the molecular docking provides advantageous alternatives wherein electrostatic and hydrophobic based docking is feasible. We performed the docking (see S1, SI text for details) of Aβ in a sequential (monomer-by-monomer) manner to form oligomers up to eicosamer. Aβ consists of 4 polar, 12 ionic, and 26 hydrophobic residues. There are 3 electrostatic patches (EP1(R5-S8), EP2(H13-K16), and EP3(S26-K28)) and 2 hydrophobic patches (HP1(L17-A21) and HP2(G29-A42)) on the SASA of the Aβ monomer (Fig. 2(a) and (b)). The electrostatic (Fig. 2(c)) (hydrophobic (Fig. 2(d))) based docking represent the case where there is a non-zero (zero) charge on the protein monomers; hence, represents the scenarios of experiments performed at pH greater/smaller than the pI (electrostatic based) of the protein and at the pI (hydrophobic based), respectively. The polar/ionic or hydrophobic patches on the oligomers should not be exhausted for the successful evolution of the oligomers into FSA (Fig. 2(e) and Fig. 2(f)). This was inspected by examining the docked structures in different orientations to assure that there remains a polar/ionic or hydrophobic patch on the SASA of the oligomer to propagate the self-assembly further (see S1, SI text).

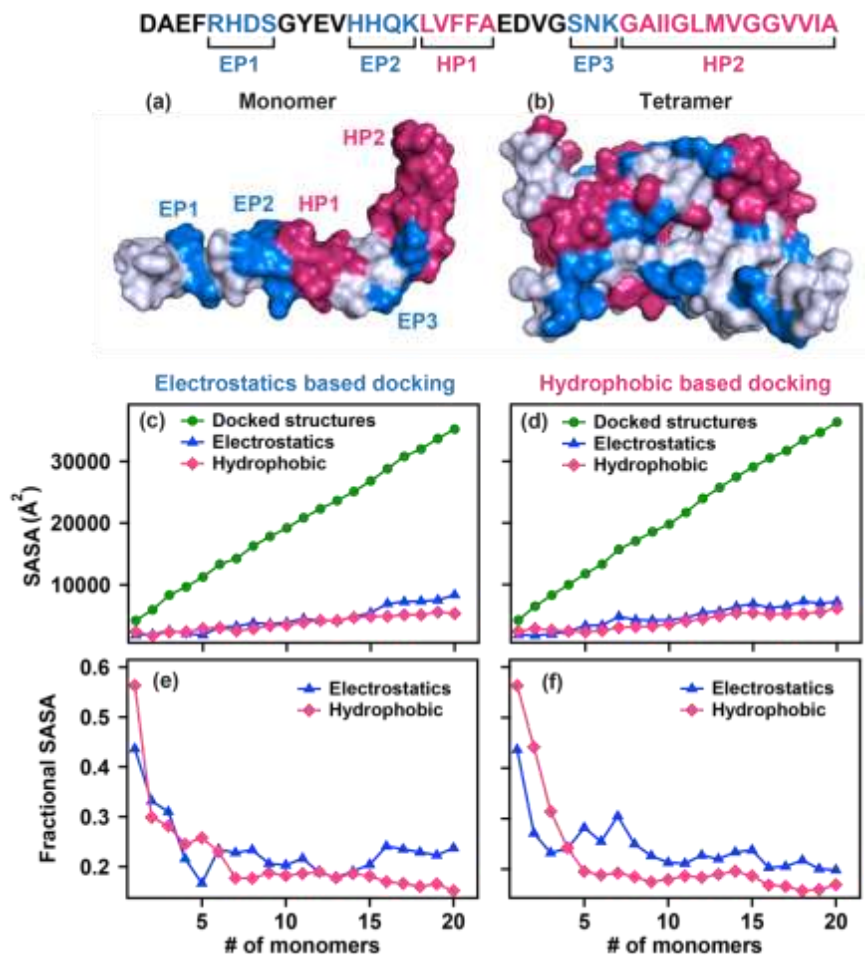

Fig. 2. The molecular docking was performed on the solution structure of Aβ (PDB ID: 1IYT). The electrostatic and hydrophobic patches on the (a) monomer and (b) docked tetramer structure. The SASA of the docked structures, polar/ionic, and hydrophobic residues on the SASA of the docked structures formed using (c) electrostatics based and (d) hydrophobic based docking, respectively. The fractional SASA of the polar/ionic and hydrophobic residues on the SASA of the docked structures formed using (e) electrostatics based and (f) hydrophobic based docking, respectively.

The change in free energy, for the formation of spherical oligomers, was used before to understand the formation of superstructures formed during protein aggregation in bulk[14]. Following a similar approach, we base our model on the calculation of the change in free energy for the formation of FSA on a substrate given by,

$$\Delta F_{frac} = -\Delta F_{oli} - \frac{(n_e E_e + n_h E_h)}{2} + \frac{3S}{R_c}\frac{4\pi R_c^3}{3} + \frac{(Nq_{eff}e)^2}{8\pi\varepsilon\varepsilon_0 R_c} - k_B T log(\frac{\varphi}{N}) \qquad (1)$$

Where $-\Delta F_{oli}$ is the change in free energy due to the formation of spherical oligomers[14], $n_e$ ($n_h$) is the number of contacts between the electrostatic-electrostatic (hydrophobic-hydrophobic) patches in the oligomers of the FSA, and $E_e$ and $E_h$ are binding energy per electrostatic and hydrophobic contact, respectively. $S$ is the surface tension due to the electrostatic and hydrophobic patches on the SASA of an oligomer which are left uncovered after an oligomer attach to the FSA, $R_c$ is the effective radius of the FSA, $q_{eff}$ is the effective number of charges on an oligomer, $e$ is the elementary charge, $\varepsilon$ is the relative dielectric constant, $\varepsilon_0$ is the permittivity in a vacuum, $N$ and $\varphi$ is the number and volume fraction of the oligomers in the FSA. The first term represents the stage1 (Fig. 1(a)) of the process, where the oligomers are formed. The second and the third terms[18] accounts for the event when an oligomer attaches to the growing FSA (stage2, Fig. 1(b)) and thus contains the information of the interactions involved due to the electrostatic and the hydrophobic patches. The fourth term is the electrostatic energy to bring an oligomer having a net charge, due to the presence of electrostatic patches on its SASA, far from the evolving FSA which account for the electrostatic repulsion[18, 19] between the oligomers. The fifth term addresses the entropic energy[20] due to the loss of a translational degree of freedom when an oligomer attach to the growing FSA.

In a protein solution, the magnitude of hydrophobic interactions, which is less compared to the electrostatic interactions, is among other interactions which may modulate the electrostatics by changing $q_{eff}$ [21-25]. The $q_{eff}$ may also be changed by changing the pH of the protein solution[9, 10, 12]. The higher $q_{eff}$ means higher repulsive interactions between the oligomers. In presence of sufficient $q_{eff}$, the oligomers may be stable and thus instead of FSA, chain-like morphologies will be observed (see S2, SI text). Also, the charges on the substrate may disturb the dynamics necessary for the self-assembly. However, during the formation of FSA, the uncovered hydrophobic patches, $n_h$, on the SASA of the oligomers would try to get covered in order to decrease the free energy. So, in an energetically favored self-assembled system, the hydrophobic patches will be hidden inside the self-assembly, and the electrostatic patches will be oriented to have minimum repulsive interaction between the electrostatic patches. The $\varphi$ of the oligomers in the FSA is the indicator of the complexity of the system, very similar to the fractal dimension ($d_f$) used to characterize the FSA.

As shown by Fodera et. al[14], the free energy of the evolving FSA (Eq. (1), terms 2-5), with increasing $R_c$, can be elegantly converted into a generic equation in terms of $d_f$ of the growing self-assembly, number of nearest neighbors to an attached oligomer $M$, the effective number of interactions $f$, and the effective radius of an oligomer $a$ to capture the complete evolution of the FSA.

$$\Delta F = -M \left(\frac{R_c}{a}\right)^{d_f} n_E k_B T + \frac{f n_E k_B T d_f}{a^{(d_f-1)}} R_c^{(d_f-1)} + \frac{(q_{eff}e)^2 R_c^{(2d_f-1)}}{8\pi\varepsilon\varepsilon_0 a^{2d_f}} - k_B T \log \frac{a^3}{R_c^3} \qquad (2)$$

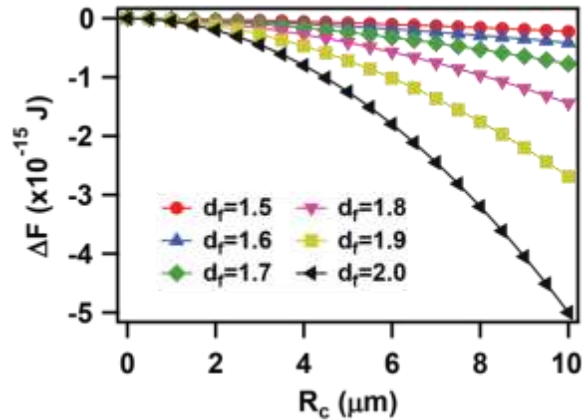

Fig. 3. The change in free energy as a function of the effective radius of the FSA computed at different $d_f$. For calculation, $a$=20 nm, $n_E k_B T$=$10^2 k_B T$, $q_{eff}e$=2e where chosen.

For the complete details of the theoretical calculation for the conversion of Eq. (1) (terms 2-5) to Eq. (2), the reader is directed to the supplementary information of Ref 14 (supplemental/10.1103/PhysRevLett.111.108105). The free energy of the evolving FSA with respect to $R_c$ depends on the complexity of the FSA (Fig. 3). Higher is the $d_f$, larger is the $\varphi$ of the oligomers in the FSA (Eq. 2). There may be different reasons for protein self-assemblies to have eventually different $d_f$. In the process of self-assembly, the stage1 (Fig. 1(a)) sets the platform for the stage2 (Fig.1 (b)). In stage1, $n_e$, $n_h$, and the fractional coverage of the patches on the SASA is decided based on the ambient conditions under which the protein solution is drop-casted, which is reflected in different aggregation pathway followed at different pH. Based on these parameters, the $q_{eff}$ is decided, and based on the other interactions defined in Eq. (1), the fractal self-assembly grow to have distinct $d_f$.

To test the implications of Eq. (1) and Eq. (2), we performed simulation on a square lattice using a modified version of the previously reported patchy diffusion-limited aggregation (DLA)[26-28]. The specific details of the simulation were inspired by the results of the molecular docking (Fig. 2). In the simulation, oligomers were represented by a collection of circular particles placed on the lattice (Fig. 4 (a)). This was done to incorporate the asymmetric geometry and rough SASA of the oligomers. During an attachment of an oligomer to the growing fractal, one or more electrostatic and hydrophobic patches may face the substrate and thus may not participate in the evolution of the self-assembly. This was addressed by choosing only 2-4 patches out of total 5 patches (electrostatic (EP1, EP2, and EP3) + hydrophobic (HP1 and HP2)), assuming the rest of the patches are either facing the substrate or are inside the interface of the formed oligomer. These patches having randomly chosen sizes were assigned randomly on the boundaries of the oligomers (Fig. 4 (b)). These oligomers were generated sufficiently far (two times of $R_c$) from the center of the lattice. The oligomer was randomly moved until it finds a hydrophobic patch on the evolving fractal to attach. Moreover, the orientations of the electrostatic patches were inspected to choose a configuration having minimum repulsive electrostatic interaction between the electrostatic patches. The previous steps were repeated to obtain the FSA. The effect of pH was incorporated by changing the effective charge, $q_{eff}$, in Eq. (1).

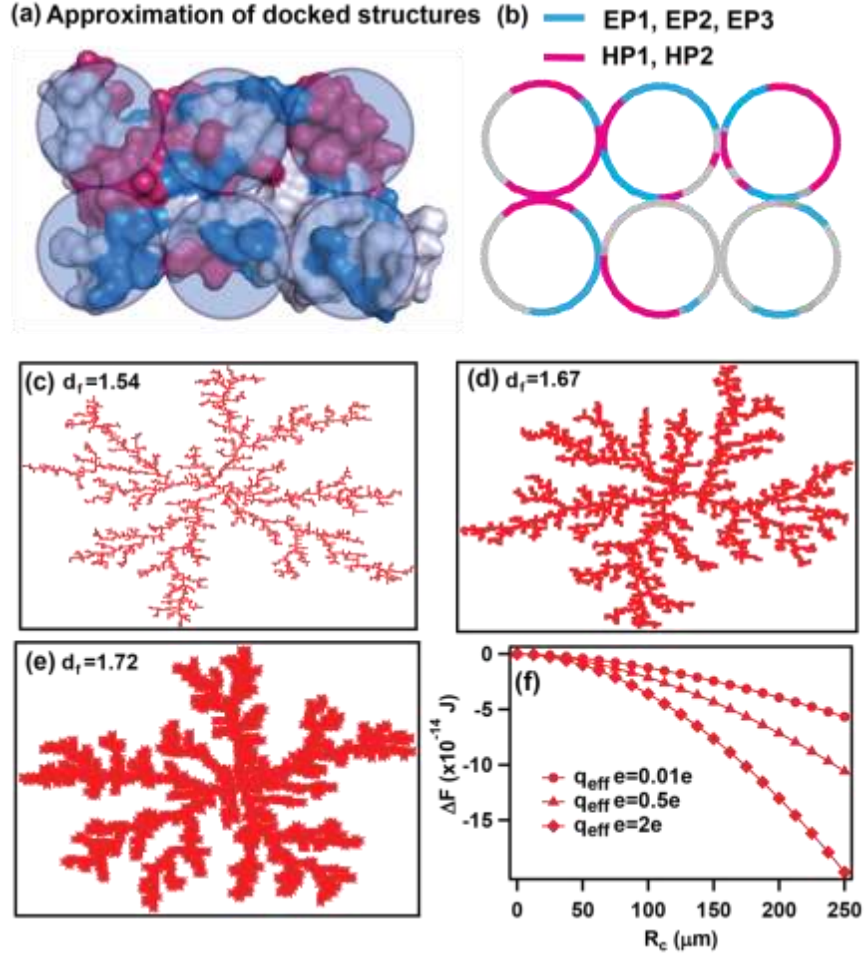

Fig. 4. Simulation protocol using patchy DLA. (a) The approximation of the docked structures using identical circles, (b) a randomly generated cluster approximating the anisotropic geometry, the electrostatic and the hydrophobic patches, and the roughness on the SASA of tetramer shown in (a). The simulated FSA obtained with (c) $q_{eff}e$=0.01e, (d) $q_{eff}e$=0.5e, and (e) $q_{eff}e$=2e. The $d_f$ of the self-assemblies were 1.54±0.02, 1.67±0.03, and 1.72±0.03, respectively. (f) The evolution of free energy as a function of $R_c$ of the simulated fractals.

Since the number ($n_e$ and $n_h$) and positions of the electrostatic and hydrophobic patches were chosen randomly in the simulation protocol, the information of stage1 (Fig. 1(a)) is not acquired (simulated) in the simulation. Therefore, we varied $q_{eff}e$ in the simulation to represent the scenarios where the net charge on the oligomers can be controlled. Fig. 4(c), (d), and (e) represents the simulated FSA for $q_{eff}e$=0.01e, $q_{eff}e$=0.5e, and $q_{eff}e$=2e, respectively. In the case of $q_{eff}e$=0.01e, there are very small electrostatic repulsions between the oligomers, and thus oligomers can attach to the growing self-assembly more quickly as there is no energy barrier that is required to overcome. This is reflected in Fig.

4(c), where thin branched structures were observed. For $q_{eff}e=2e$, the electrostatic repulsive interactions partially stabilizes the oligomers, and thus for the fractal self-assembly, the oligomers have to overcome the repulsive interactions. The self-assembly thus formed will be compact as the oligomers have to find appropriate orientations to minimize the free energy. The increase in the $q_{eff}e$ increases the repulsive interaction and inhibits the smooth propagation of the FSA but gives rise to compact fractal structures, as observed in Fig. 4(d) and Fig. 4(e). The evolution of the free energy for these three FSA was computed with the $R_c$ of the simulated fractals. The fractal with higher $d_f$ showed greater steepness in the evolution of the change in free energy. This is consistent with the theoretical evolution of $\Delta F$ with respect to $R_c$ obtained with Eq. (2), shown in Fig. 3.

Further, to correlate the theoretical results and the simulated fractals with experimental observations, the FSA of Aβ were recorded using a confocal microscope (Fig. 5). The Aβ protein solution was drop-casted on glass slides to observe the fractal self-assembly of Aβ at three different pH (see S3, SI text for experimental details). The concentration of the protein solution, at which the FSA was observed, was optimized, and the experiment was performed with the protein solution prepared in two different media- water and PBS buffer. We first confirmed that the self-assemblies were from the protein and their aggregates and not from the salt present in the solution. For Aβ solution prepared in PBS buffer, we used a scanning electron microscope (SEM) with energy dispersion X-ray (EDAX) analysis, where the significant content of nitrogen in the self-assemblies was adopted as a parameter of confirmation for the presence of protein and their aggregates in the self-assemblies (see S4(a), SI text). For Aβ solution prepared in water, the confocal microscopy was used (see S4(b) SI text), wherein the fact that no fluorescence signal can be obtained unless a fluorophore (here ThT) binds to the β-sheet structures present in the oligomers was employed to confirm the presence of Aβ in the FSA (Fig. 5).

To elucidate the role of electrostatic interactions, the pH of the Aβ solution was varied. At acidic pH (3.5±0.1), the dendritic structure was observed. At physiological pH (7.5±0.1), thin branched fractals, and at basic pH (11.5±0.1), thicker branches compared to that of the physiological pH were observed (Fig. 5). It is to be noted that the pI of the Aβ peptide solution is 6.67, which means that at pH 7.5±0.1, there will small $q_{eff}e$ on the oligomers. Thus at this pH, thin branched fractals are expected, which is consistent with simulated fractal obtained for $q_{eff}e=0.01e$ (Fig. 4(c)). At basic pH, there is a net negative charge on the proteins, indicating protein oligomers have to cross the barrier, resulting in thick branched, compact fractal structures. At pH 3.5±0.1, there is a net positive charge but smaller in magnitude compared to pH 11.5±0.1. Thus, dendrite-like fractal morphologies were observed at pH 3.5±0.1. These experimental observations are in agreement with the simulated fractals observed at different strengths of electrostatic repulsive interactions (Fig. 4(c), (d), and (e)). Moreover, the $d_f$ of the FSA through experiments at pH 3.5±0.1, 7.5±0.1, and 11.5±0.1 were 1.68±0.04, 1.54±0.03, and 1.81±0.04, respectively. This is consistent with the predictions of the model and the simulation results where FSA with higher $d_f$ are expected to have compact structures.

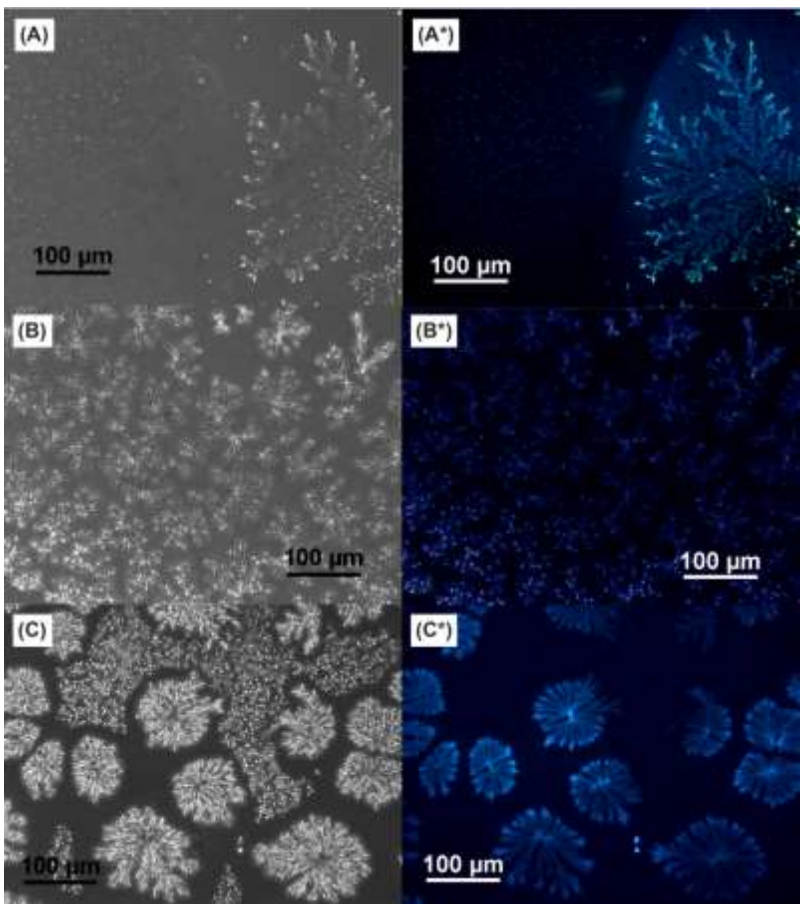

Fig. 5. The non-confocal and the confocal mode images of the FSA of Aβ obtained with Aβ solutions prepared in water at (A) and (A*) pH 3.5±0.1; (B) and (B*) pH 7.5±0.1; (C) and (C*) pH 11.5±0.1, respectively.

It is interesting to note an attractive feature in the FSA at pH 7.5±0.1 and pH 11.5±0.1. At pH 7.5±0.1, there is an indication of the central point of the fractal to act as a bifurcation point for the morphology. The bifurcation characteristics became even more apparent at pH 11.5±0.1. This indicates the presence of other parameters, the interplay between which affects the shape plasticity of the self-assembly[29]. The previous results also suggest that the surface thermodynamics and the crystal morphologies are related[30]. The detailed study of the presence of a bifurcation point in these FSA is the subject matter of our on-going investigation.

In summary, using theoretical modeling, simulations with a modified algorithm of the patchy DLA, and qualitative comparison of these results with the experimentally recorded FSA of Aβ, we show that the FSA of protein are mediated through the electrostatic and the hydrophobic patches present on the SASA of the protein. The FSA shows distinct morphologies governed by the repulsive electrostatic interactions, which can be modulated by changing the pH of the protein solution in experiments. The model can help understand the physicochemical interactions, which may be needed to design appropriate protein self-assemblies for desired technological applications especially in biomedical[31] and construction of nanodevices[32].

We thank Somnath Maji from the Department of Biotechnology, IIT Kharagpur, for his assistance in acquiring the confocal microscopic images of the Aβ FSA. AS and SK are thankful to the IIT Kharagpur for FESEM imaging facility and MHRD for financial assistance. This work is supported by the Department of Science and Technology, India, Grant number: CRG/2019/001684, project code: KML.